\documentclass[12pt]{article}
\usepackage{amsmath,amsfonts,amssymb,latexsym}
\usepackage{epsfig}
\usepackage{graphicx}

\newcommand{\be}{\begin{equation}}
\newcommand{\ee}{\end{equation}}

\newcommand{\ds}{\displaystyle}

\def\C{\mathbb{C}}

\def\H{\mathcal{H}}

\def\Z{\mathbb{Z}}

\begin{document}

\thispagestyle{empty}
\hfill \today

\vspace{2.5cm}

\begin{center}
\bf{\LARGE
Lie algebra representations \\[0.3cm] 
and \\[0.4cm]
 rigged Hilbert spaces: the SO(2)  case}
\end{center}

\bigskip\bigskip

\begin{center}
Enrico Celeghini$^1,2$, Manuel Gadella$^2$ and Mariano A del Olmo$^2$
\end{center}

\begin{center}
$^1${\sl Dipartimento di Fisica, Universit\`a  di Firenze and
INFN--Sezione di
Firenze \\
I50019 Sesto Fiorentino,  Firenze, Italy}\\
\medskip

$^2${\sl Departamento de F\'{\i}sica Te\'orica and IMUVA, Universidad de
Valladolid, \\
E-47005, Valladolid, Spain.}\\
\medskip

{e-mail: celeghini@fi.infn.it, mgadella@gmail.com, marianoantonio.olmo@uva.es}

\end{center}

\begin{abstract}
It is well known that related with the irreducible representations of the Lie group $SO(2)$ we find a discrete basis as well a continuous one. In this paper we revisited this situation under the light of  Rigged Hilbert spaces, which are the suitable framework to deal  with both discrete and bases in the same context and in  relation with physical applications.
\end{abstract}


%%%%%%%%%%%%%%%%%%%%%%%%%%%%%%%%%%%%%%%%
%%%%%%%%%%%%%%%%%%%%%%%%%%%%%%%%%%%%%%%%

\section{Introduction}

In the last years we have been involved in a program of revision of the  connection between  special functions  (in particular, 
Classical Orthogonal Polynomials), Lie groups, differential equations and physical spaces.
We have obtained the ladder algebraic structure for different orthogonal polynomials, like Hermite, Legendre, Laguerre  \cite{EM}, Associated Laguerre Polynomials, Spherical Harmonics, etc. \cite{EMa,EMM}.
In all cases, we have obtained a symmetry group. The corresponding orthogonal polynomial is associated to a particular representation  of  its Lie group.
For instance, for the Associated Laguerre Polynomials and the Spherical Harmonics, we obtain the symmetry group $SO(3,2)$ and in both cases they support a unitary irreducible representation (UIR) with quadratic Casimir $-5/4$.
Both are bases of square integrable functions defined on $(-1,1)\times \Z$ and on the sphere $S^2$, respectively.
In any case we get {discrete basis}  and continuous basis. 

On the other hand,  the Rigged Hilbert Space (RHS) is a suitable  framework  for a description of quantum states, when the use of both discrete basis, i.e.,  complete orthonormal sets, and generalized continuous basis like those used in the Dirac formalism are necessary
\cite{GEL}--\cite{GG2}.
As  mentioned above,
this is a typical situation arisen when we deal with special functions, which hold  discrete  labels and depend on continuous variables. 
We have analysed this situation for the  Hermite and Laguerre functions motivated also by possible applications on signal theory in recent papers  \cite{EM2015,EMG}.  Moreover, the RHS fit very well with Lie groups \cite{MAURIN1} and also with semigroups, see \cite{GGRW} and references therein.

In this paper, we continue the study of the relation between  Lie algebras, special functions and RHS. Here we propose a revision of the simplest case provided by the Lie algebra $so(2)$ related to the Fourier series, both for its practical interest and as an introduction to our ideas. 
Since  continuous and discrete bases are involved \cite{WKT}, the framework of the RHS is the appropriate one for this case too. Thus, we use the $so(2)$ Lie algebra and its UIR   as to construct the RHS associated to these representations.

%%%%%%%%%%%%%%%%%%%%%%%%%%%%%%%%%%%%%%%%%%%%%%

\section{Rigged Hilbert Spaces}

There are several reasons to assert that Hilbert spaces are not sufficient for a thoroughly formulation of QM even within the non-relativistic context. We can mention, for instance, the Dirac formulation \cite{DIRAC} where   operators with  continuous spectrum play a crucial role  (see also \cite{GG1} and references therein) and their eigenvectors are not in the Hilbert space of square integrable wave functions.
 Another example is related with the proper definition of Gamow vectors \cite{BG}, which are widely used in calculations including unstable quantum systems and are non-normalizable.  We can also  refer to  formulations of time asymmetry in QM   that may require the use of tools  more general than Hilbert spaces \cite{BGK}.

The proper framework that includes naturally the Hilbert space and its features, which are widely used in QM,   is the RHS. 

The Rigged Hilbert Spaces were introduced by Gel'fand and collaborators \cite{GEL} in connection with the spectral theory of self-adjoint operators. They also proved, together with Maurin \cite{MAU}, the nuclear spectral theorem \cite{GG1,GG2}.
The RHS formulation of Quantum Mechanics (QM) was introduced by Bohm and Roberts around 1965 \cite{RO,B}.

A Rigged Hilbert Space  (also called Gelf'and triplet)
 is a triplet of spaces  
\[
\Phi\subset\mathcal H\subset \Phi^\times\, ,
\]
with
$\mathcal H$   an infinite dimensional separable Hilbert space, $\Phi$ (test vectors space)    a dense subspace of $\mathcal H$ endowed with its own topology, and
$\Phi^\times$ is the dual /antidual space of $\Phi$. 

The  topology considered on $\Phi$ is finer  (contains more open sets) than the topology that 
$\Phi$ has as subspace of $\mathcal H$, and
$\Phi^\times$ is equipped with a topology compatible with the dual pair
 $(\Phi,\Phi^\times)$ \cite{HOR}, usually the weak topology.
One  consequence of the topology of $\Phi$ \cite{PI,GG1}  is that all sequences which converge on $\Phi$, also converge on $\mathcal H$, the converse being not true. 
The difference between topologies gives rise that the dual space of $\Phi$, $\Phi^\times$, is bigger than $\mathcal H$, which is self-dual.

Here, we shall consider the dual $\Phi^\times$ of $\Phi$, i.e.,  
 any $F\in\Phi^\times$ is a continuous linear mapping from $\Phi$ into $\C$.
Linearity and antilinearity mean, respectively,  that for any pair of vectors $\psi,\varphi\in\Phi$, any pair
$F,G \in\Phi^\times$ and any pair $\alpha,\beta \in \C$ we have 
\[\begin{array}{l}
\langle F|\alpha\psi+\beta\varphi\rangle=\alpha\,\langle F|\psi\rangle+\beta\,\langle F|\varphi\rangle
\,,\\[0.4cm]
\langle \alpha F+\beta G | \psi\rangle=\alpha^* \,\langle F|\psi\rangle+\beta^*\,\langle G |\psi\rangle\,,
\end{array}
\]
 where we have followed the Dirac bra-ket notation and the star denotes complex conjugation.

A crucial property to be taken under consideration is that 
if   $A$ is a densely defined operator on $\mathcal H$, such that $\Phi$ be a subspace of its domain and that 
$ A\varphi\in\Phi $ for all $\varphi\in\Phi$,
  we say that $\Phi$ reduces $A$ or that $\Phi$ is invariant under the action of $A$, (i.e., $A\Phi\subset\Phi$).  
In this case, $A$ may be extended unambiguously to the dual $\Phi^\times$ by making use of the duality formula
\be\label{dualidad}
\langle A^\times\,F|\varphi\rangle :=\langle F|A\varphi\rangle\,,\qquad \forall\,\varphi\in\Phi\,, \;\;\forall\,F\in\Phi^\times\,.
\ee
If $A$ is continuous on $\Phi$, then $A^\times$  is continuous on $\Phi^\times$.

The  topology on $\Phi$ is given by an infinite countable set of norms
$\{||-||_{n=1}^\infty\}$.
A linear operator $A$ on $\Phi$ is continuous  if and only if for each norm $||-||_n$ there is a  $K_n>0$ and a finite sequence of norms $||-||_{p_1}, ||-||_{p_2}, \dots, ||-||_{p_r}$ such that for any $\varphi\in\Phi$, one has \cite{RS}
\begin{equation}\label{continuidad}
||A\varphi||_n\le K_n\,\left(||\varphi||_{p_1}+||\varphi||_{p_2}+\dots+||\varphi||_{p_r}\right)\,.
\end{equation}
The same result applies to check the continuity of any linear or antilinear mapping 
$
F:\Phi\longmapsto\mathbb C .
$
 In this case, the norm $||A\varphi||_p$  should be replaced by the modulus $|F(\varphi)|$.

%%%%%%%%%%%%%%%%%%%%%%%%%%%%%%%%%%%%%%%%
%%%%%%%%%%%%%%%%%%%%%%%%%%%%%%%%%%%%%%%%

\section{A paradigmatic case: RHS for $SO(2)$}

As mentioned before, we have considered   the most elementary situation provided by $SO(2)$, where we have  two  RHS serving as  support of unitary equivalent representations of $SO(2)$. 
One of these RHS is a  concrete RHS constructed with functions or generalised functions and the other one is an  abstract RHS. 
A mapping  of the test vectors of the abstract RHS   gives the  test functions of the concrete one.
 Also we have to  adjust the topologies so that the elements of the Lie algebra be continuous operators on both test  spaces and their corresponding duals.

Let us remember that $SO(2)$   is the group of rotations on the Euclidean plane.
It is a one-dimensional (1D) abelian Lie group, parametrized by $\phi\in[0,2\pi)$. 
 
The  elements $R(\phi)$ of $SO(2)$ satisfy the  product law
\[
R(\phi_1)\cdot R(\phi_2)=R(\phi_1+\phi_2)\,, \qquad {\rm mod }\; 2\pi\,.
\]
 
Here, we are considering  two equivalent families of UIR of $SO(2)$:
 one of them  supported by the Hilbert space $L^2[0,2\pi]$ (via the regular representation, that contains once  all the UIR, each one related to an integer number) and 
another set of UIR (also labelled by $\mathbb Z$) supported by an abstract 
infinite dimensional separable Hilbert space $\mathcal H$. 

%%%%%%%%%%%%%%%%%%%
%%%%%%%%%%%%%%%%%%%
\subsection{UIR  supported by the HS $L^2[0,2\pi]$}

We consider the UIR characterised by the unitary operator on $L^2[0,2\pi]$
\be\label{um}
\mathcal U_m(\phi):= e^{-im\phi}\,, \;\; \forall\,\phi\in[0,2\pi)\,,\;   m\in\mathbb Z \;\text{(fixed)}.
\ee
An   orthonormal basis for $L^2[0,2\pi]$ is given by the
 sequence of functions $\phi_m$
 labelled by   $m\in\mathbb Z$
\[
\phi_m\equiv \frac1{\sqrt{2\pi}}\, e^{-im\phi}\,, \qquad m\in\mathbb Z.
\]
Thus, any Lebesgue square integrable function $f(\phi)$ of $ L^2[0,2\pi]$ can be written as
\begin{equation}\label{fphi}
\ds f(\phi)=\sum_{m=-\infty}^\infty f_m\,\phi_m\,,\ee
with 
\be\label{coefficients}
 \ds f_m=\frac1{\sqrt{2\pi}} \int_0^{2\pi} e^{im\phi}\,f(\phi)\,d\phi,
\end{equation}
under the condition that 
\[
  \sum_{m=-\infty}^\infty |f_m|^2=\int_0^{2\pi} |f(\phi)|^2 \,d\phi<+\infty\,.
  \]
Note that the complex numbers $f_m$ are the Fourier coefficients of  $f(\phi)$. 

The functions  $\mathcal U_m= e^{-im\phi}$ satisfy   the following orthogonality and completeness relations:
\[\begin{array}{l}
\ds \frac1{2\pi} \int_0^{2\pi} \mathcal U^\dagger_m (\phi)\,\mathcal U_n(\phi)\,d\phi=\delta_{m,n}\,,\\[0.4cm] \ds
 \frac1{2\pi} \sum_{m=-\infty}^\infty \mathcal U^\dagger_m (\phi)\,\mathcal U_m(\phi')=\delta(\phi-\phi')\,.
\end{array}\]

%%%%%%%%%%%%%%%%%%%%%
%%%%%%%%%%%%%%%%%%%%%
\subsection{UIR  on an 
infinite-D separable HS}

 Equivalently, we may construct another set of UIR's of $SO(2)$ labelled by $\mathbb Z$ and supported on an abstract infinite dimensional separable Hilbert space $\mathcal H$.

Let  $\{|m\rangle\}_{m\in\Z}$ be an  orthonormal basis of $\mathcal H$.
There is a unique natural  unitary mapping $S$ such that
\[\begin{array}{cccr}
 \mathcal H&\stackrel{S}\longmapsto &L^2[0,2\pi]&\\[0.3cm]
|m\rangle &\longmapsto & S|m\rangle=\phi_m,&\qquad \forall  m\in\Z\,.
\end{array}\] 
Let us consider the subspace $\Phi$ of $\mathcal H$
of vectors  
\begin{equation}\label{fm}
|f\rangle=\sum_{m=-\infty}^\infty a_m\,|m\rangle \in\mathcal H\,,\qquad a_m\in \C\,,
\end{equation}
such that 
\be\label{modulof} 
\langle f|f\rangle_p\equiv ||f||_{ p}^2:=\sum_{m=-\infty}^\infty |a_m|^2\,{ |m+i|^{2p}}<\infty\,,
\ee
for any $p=0,1,2,\dots$
Since 
$\Phi$ contains all finite linear combinations of the basis vectors $|m\rangle$ is dense on $\mathcal H$.
We endow $\Phi$ with the metrizable topology generated by the norms 
$
 ||f||_p,\; ( p=0,1,2,\dots).
$
  In this way we have constructed a RHS: 
$ 
  \Phi\subset \mathcal H\subset \Phi^\times.
  $

  Considering that the unitary mapping  $S$ transports the topologies, we get two RHS
\[\begin{array}{ccccc}
\Phi&\subset &\mathcal H&\subset& \Phi^\times\\[0.4cm]
 (S\Phi)&\subset& L^2[0,2\pi]&\subset &(S\Phi)^\times
\end{array}
\]
such that 
  $\Phi$ and $\mathcal H$ have the  discrete basis $\{|m\rangle\}_{m\in\Z}$ and 
$(S\Phi)$ and $L^2[0,2\pi]$ have 
  its equivalent discrete basis   $\{\phi_m\}_{m\in\Z}$. 
Now, we may define continuous basis in both RHS as follows.   
Since these two RHS are unitarily equivalent, it is enough to construct the continuous basis on the abstract RHS   and to induce the equivalent one in the other RHS.

 Let us consider  the abstract RHS
$
 \Phi \subset \mathcal H \subset \Phi^\times.$ 
 Since  $|m\rangle \in \H$ we can consider  $\langle m |\in \H^\times=\H$. Then, for any 
 $\phi\in [0,2\pi)$, we can  define a ket $|\phi\rangle$ such that 
\[
\langle m|\phi\rangle:= \frac1{\sqrt{2\pi}}\, e^{im\phi}.
\]
 From the duality relation $\langle\phi|m\rangle=\langle m|\phi\rangle^*$  and for any $
|f\rangle=\sum_{m=-\infty}^\infty a_m\,|m\rangle\;\in\Phi$ we get
\[
\langle\phi|f\rangle= \sum_{m=-\infty}^\infty a_m\,\langle \phi|m\rangle= 
 \frac1{\sqrt{2\pi}}\ \sum_{m=-\infty}^\infty a_m\, e^{-im\phi}\,,
\]
where  $a_m=f_m$ as in \eqref{fphi}. 
The action of $\langle\phi|$ on $\Phi$, $ \langle\phi|f\rangle$,   is well defined
since 
the following  series is absolutely convergent
\begin{equation*}\label{8}\begin{array}{l}\ds
\sum_{m=-\infty}^\infty |a_m|= \sum_{m=-\infty}^\infty \frac{ |a_m|\,{ |m+i|}}{|m+i|}\\[0.4cm]
\quad\ds \le \sqrt{\sum_{m=-\infty}^\infty |a_m|^2\,{|m+i|^2}} \sqrt{\sum_{m=-\infty}^\infty \frac1{ |m+i|^2}}\,.
\end{array}
\end{equation*}
Note that both series in the second row of this inequality  converge: the first one    because   it verifies eq.  \eqref{modulof}  for $ p=1$, and it is obvious for  the second series. 

Since ${ |\langle\phi|f\rangle|\le C\,||f||_1}$ with
\[
 \begin{array}{l}\ds
||f||_1=\sqrt{\sum_{m=-\infty}^\infty |a_m|^2\,|m+i|^2}\,,\\[0.6cm]
\ds C=\sqrt{\sum_{m=-\infty}^\infty \frac1{|m+i|^2}}\,,
\end{array}
 \] 
 then $ \langle \phi |\in\Phi^\times$.
Note that $\langle\phi|f\rangle=\langle f|\phi\rangle^*$, hence $\langle\phi|$ is an antilinear map on 
$\Phi$, while $|\phi\rangle$ is linear. 

On the other hand,  
$\{|\phi\rangle\}_{\phi\in[0,2\pi)}$, is a continuous basis. In fact, if we apply the map $S$ to an arbitrary  $ |f\rangle\in \Phi$ as in \eqref{fm}, we obtain that $S|f\rangle\in S \Phi\subset L^2[0,2\pi]$ and
\be\begin{array}{ll}\label{sf}
 S|f\rangle&=\displaystyle\sum_{m=-\infty}^\infty a_m\, S|m\rangle
 =\sum_{m=-\infty}^\infty a_m\,\frac{e^{-im\phi}}{\sqrt{2\pi}} \\[0.5cm]
& =\langle\phi|f\rangle=f(\phi)\,.
 \end{array}
\ee
If  $|f\rangle,|g\rangle\in \Phi$, then $f(\phi)=S|f\rangle$ and $g(\phi)=S|g\rangle$ belong to $(S\Phi)\subset L^2[0,2\pi]$. Thus, 
and due to the unitarity of $S$, we get
\be\begin{array}{lll}\label{pescalar}
\langle f|g\rangle&=&\ds\int_0^{2\pi} f^*(\phi)\,g(\phi)\,d\phi\\[0.3cm]
&=&\ds \int_0^{2\pi} \langle f|\phi\rangle\langle \phi| g\rangle\,d\phi \,,
\end{array}
\ee
and thus
\be\label{iidentidad}
\mathbb I=\int_0^{2\pi} |\phi\rangle\langle \phi|\,d\phi\,.
\ee
Applying this identity to  $|f\rangle\in\Phi$, we have 
\be\label{masidentidad}
\mathbb I |f\rangle= \int_0^{2\pi} |\phi\rangle\langle \phi|f\rangle\,d\phi = \int_0^{2\pi} f(\phi)\,|\phi\rangle\,d\phi\,.
\ee
This gives a  span of $|f\rangle$ in terms of $|\phi\rangle$ with coefficients $f(\phi)$ for all $\phi\in [0,2\pi)$, which shows that 
$\{|\phi\rangle\}$ is a continuous basis on $\Phi$, although its  elements are not in $\Phi$ but  instead in
$\Phi^\times$. Since  
$\langle \phi |$ acts on  $\Phi$ only (not on all  $\mathcal H$), then for an arbitrary $|g\rangle\in\Phi$ we have
\[
\langle g|\mathbb I f\rangle= \int_0^{2\pi} \langle g|\phi\rangle\langle \phi|f\rangle\,d\phi =\langle g|f\rangle\,.
\]
Because of the definition of RHS to any $|f\rangle\in\Phi$ corresponds a $\langle f|\in\Phi^\times$ and the action of $\langle f|$ on any $|g\rangle\in \Phi$   is given by the scalar product  $\langle f|g\rangle$ \eqref{pescalar} from $L^2[0,2\pi]$. 
Thus,   $\mathbb I$ is the canonical injection $\mathbb I:\Phi\longmapsto\Phi^\times$, and it is continuous. 
Moreover, $|\phi\rangle$ is a continuous linear functional on $\Phi$  then $\langle \phi|$ is a continuous antilinear functional on $\Phi$. Consequently,  
\[
f(\phi)=\langle\phi|f\rangle =\int_0^{2\pi} \langle \phi|\phi'\rangle\langle\phi'|f\rangle\,d\phi'\,,
\]
and then,
\be \label{interno}
\langle \phi|\phi'\rangle=\delta(\phi-\phi').
\ee
Therefore,
 the set  $\{|\phi\rangle\}$ satisfies the relations of orthogonality \eqref{interno} and completeness \eqref{iidentidad}
 that allow us to write
 \be\label{phibases}
\mathbb I |f\rangle= \int_0^{2\pi} |\phi\rangle\langle \phi|f\rangle\,d\phi = \int_0^{2\pi} f(\phi)\,|\phi\rangle\,d\phi\,,
\ee
 showing, once more,  that $\{|\phi\rangle\}$ is a continuous basis. 
 Note that $\{|\phi\rangle\}$ spans $\Phi$ and not $\mathcal H$.

There are some formal relations between both basis, $\{|m\rangle\}$ and 
$\{|\phi\rangle\}$. For instance,  replacing $f$ by $|m\rangle$ in \eqref{phibases} we have that
\[
|m\rangle=\int_0^{2\pi} \langle \phi|m\rangle\,|\phi\rangle\,d\phi=\frac1{\sqrt{2\pi}} \int_0^{2\pi} e^{-im\phi} \,|\phi\rangle\,d\phi\,.
\]
Since $\{|m\rangle\}$ is a basis in $\mathcal H$, the following completeness relation holds
\be\label{iiidentidad}
\sum_{m=-\infty}^\infty |m\rangle\langle m|=I\,,
\ee
where $I$ is the identity on $\mathcal H$ and also on $\Phi$. 
Do not confuse this identity with $\mathbb I$ previously defined \eqref{iidentidad}. 

Because $|m\rangle\in\Phi$, we may apply to it any element of $\Phi^\times$ so that $I$ becomes a well defined identity on the dual $\Phi^\times$  
\[
\langle \phi |\, I=\langle\phi|=\sum_{m=-\infty}^\infty \langle \phi |m\rangle\langle m| = \frac1{\sqrt{2\pi}} \sum_{m=-\infty}^\infty 
e^{-im\phi}\,\langle m |\,,
\]
which gives the second formal identity \eqref{iiidentidad}. Nevertheless and due to the absolute convergence of the series 
\[
\langle\phi|f\rangle= \sum_{m=-\infty}^\infty a_m\,\langle \phi|m\rangle=  \frac1{\sqrt{2\pi}}\ \sum_{m=-\infty}^\infty a_m\, e^{-im\phi}\,,
\]
it is easy to prove that $\langle \phi |\,I$ converges in the weak topology on $\Phi^\times$.

\subsection{Action of $so(2)$ on the RHS's}

The Hilbert space $L^2[0,2\pi]$ also supports  the regular representation of $SO(2)$, $\mathcal R(\theta)$, defined by
\[
\left[\mathcal R(\theta)\,f\right](\phi):=f(\phi-\theta)\,,\;\;\;  {\rm mod }\; 2\pi\,,\;\; \forall f\in L^2[0,2\pi], 
\]
for any 
$\theta
\in [0,2\pi)$. 

The unitary map 
$
S:\mathcal H\longmapsto L^2[0,2\pi]
$
 also allows us to transport  
 $\mathcal R$ to an equivalent representation $R$ supported on $\mathcal H$ by
\[
R(\theta)=S^{-1}\,\mathcal R(\theta)\,S\,,
 \]
such that $R(\theta)\Phi=\Phi,\;\forall \theta\in [0,2\pi)$.
Since $\mathcal R(\theta)$ is unitary on $L^2[0,2\pi]$ for any value of $\theta$ then 
$R(\theta)$ is also unitary on $\mathcal H$ due to the unitarity of $S$. 

Unitary operators leaving $\Phi$ invariant can be extended to $\Phi^\times$ by the duality formula \eqref{dualidad}, i.e., 
\[
 \langle R(\theta)F | f\rangle=\langle F | R(-\theta) f\rangle\,,\qquad \forall\,|f\rangle\in\Phi\,,\quad \forall\,F\in\Phi^\times\,.
\]
Therefore, 
\[
\langle R(\theta)\phi | f\rangle=\left[R(-\theta)\,f\right](\phi)=f(\phi+\theta)=\langle \phi+\theta |f\rangle\,.
\]
Combining both expressions  and dropping the arbitrary $| f\rangle\in\Phi$ we arrive to
\[
\langle R(\theta) \phi |\equiv \langle \phi |R(\theta) =\langle \phi+\theta |\,,\quad 
{\rm mod}\;2\pi\,,
\]
which is a rigorous expression in $\Phi^\times$.
In fact, let  $\ds| f\rangle 
\in \Phi$ as in \eqref{fm}. Then, we have
\[\begin{array}{l}\ds
{ R(\theta) \sum_{m=-\infty}^\infty a_m\,|m\rangle}= \ds S^{-1} \sum_{m=-\infty}^\infty a_m \, S\,R(\theta)\,S^{-1}\,S |m\rangle \\[0.4cm] 
\qquad =\ds S^{-1} \sum_{m=-\infty}^\infty a_m \,\mathcal R(\theta)\,\frac1{\sqrt{2\pi}}\,e^{-im\phi} \\[0.4cm] 
\qquad =\ds S^{-1}\sum_{m=-\infty}^\infty a_m\, \frac1{\sqrt{2\pi}}\,e^{-im(\phi-\theta)} \\[0.4cm] 
\qquad =\ds S^{-1} \sum_{m=-\infty}^\infty a_m\, e^{im\theta}\,\frac1{\sqrt{2\pi}}\,e^{-im\phi}\\[0.4cm] 
\qquad =\ds { \sum_{m=-\infty}^\infty a_m \, e^{im\theta} \,|m\rangle}\,\in \Phi\,.
\end{array}\]
Hence, we see that
$
R(\theta)\Phi\subset\Phi$ and since $R^{-1}(\theta)=R(-\theta)$ then 
$\Phi\subset R(-\theta)\Phi$. So,  $\Phi= R(\theta)\Phi \,,\;\; \forall\theta.
$   

The UIR' s, $U_m$, on $\mathcal H$ are given in terms of the ${\cal U}_m$, defined in \eqref{um},  by 
\[ 
U_m(\phi):= S^{-1}\mathcal U_m(\phi)\,S\,.
\]  
Let $J$ be the infinitesimal generator associated to this representation, i.e. 
$U_m(\phi)=e^{-iJ\phi}\,.$ Since $U_m$ is unitary then $J$   is self-adjoint. Its action on the vectors $|m\rangle$ is 
 \[
 J|m\rangle=m\,|m\rangle\,.
\] 
Hence  for any $| f\rangle\in \Phi$ as in \eqref{fm}, we have that
\[ 
J| f\rangle= \sum_{m=-\infty}^\infty a_m \,m \,|m\rangle\,.
\]
From the set of norms $||Jf||_{ p}^2$,  $p=0,1,2,\dots$, that we have defined in \eqref{modulof}, we obtain the following inequality
\begin{equation*}\label{15}
\sum_{m=-\infty}^\infty |a_m|^2\,m^2\,{ |m+i|^{2p}}\le \sum_{m=-\infty}^\infty |a_m|^2\,{ |m+i|^{2p+2}}\,,
\end{equation*}
which shows that $ J\Phi\subset\Phi$.
Also, this inequality   
may be also read as
\[
||Jf||_p^2\le ||f||_{p+1}\,,\;\; \forall | f\rangle\in\Phi\,,\;\; \forall p\,.
\]
Thus, we have proved that $J $ is  continuous on $\Phi$.
Moreover, since the 
self-adjoint operator $J$ verifies $J\Phi\subset\Phi$, it can be extended to $\Phi^\times$ using the duality form, i.e.,
\[
\langle JF|f\rangle=\langle F |J f\rangle\,,\qquad \forall\,|f\rangle\in \Phi\,,\quad \forall\,F\in\Phi^\times\,.
\]
Furthermore, since $J$ is continuous on $\Phi$, this extension is weakly (with the weak topology) continuous on $\Phi^\times$. In fact, since the series
\[
I|\phi\rangle=|\phi\rangle=\sum_{m=-\infty}^\infty |m\rangle\langle m|\phi\rangle = \frac1{\sqrt{2\pi}} \sum_{m=-\infty}^\infty e^{im\phi}\,|m\rangle
\] 
is weakly convergent, then  
\[\begin{array}{ll}
J|\phi\rangle&= \ds\sum_{m=-\infty}^\infty e^{-im\phi}\,J|m\rangle = \sum_{m=-\infty}^\infty e^{-im\phi}\, m \,|m\rangle\\[0.4cm]
& =\ds i\,D_\phi\,|\phi\rangle\,,
\end{array}
\]
where the  operator $D_\phi$ is defined as follows: for any $| f\rangle$ in $\Phi$ we know that 
$S| f\rangle=f(\phi)\in (S\Phi)$ as in \eqref{sf}. 
Then, 
\be\begin{array}{ll}\label{derivative}
\ds 
i\frac{d}{d\phi}\, f(\phi)&=\ds  i\frac{d}{d\phi}\,\sum_{m=-\infty}^\infty a_m\,e^{-im\phi}
\\[0.4cm]
& =\ds\sum_{m=-\infty}^\infty a_m\,m\,e^{-im\phi}\,.
\end{array}\ee
We easily conclude that the operator 
$
i\,{d}/{d\phi}
$
is continuous on $S\Phi$ with the topology transported by $S$ from $\Phi$ (norms on $S\Phi$ look like exactly as the norms on $\Phi$). 
Hence, 
\[
iD_\phi:=S^{-1}\,i\frac{d}{d\phi}\,S\,.
\]
 This definition implies that $iD_\phi$ is   continuous  on $\Phi$. 
 Moreover, it is self-adjoint on $\mathcal H$, so that it can be extended to a weakly continuous operator on $\Phi^\times$ as  the last identity in \eqref{derivative} shows.
 Therefore
 \[
 J\equiv iD_\phi
 \]
 on $\Phi^\times$.
%%%%%%%%%%%%%%%%%%%%%%%%%%%%%%%%%%%%%%%%%%%%%%%%
%%%%%%%%%%%%%%%%%%%%%%%%%%%%%%%%%%%%%%%%%%%%%%%%

\section{Conclusions} 

We have construct two RHS that support the UIR of the Lie group $SO(2)$, 
$\Phi\subset \mathcal H\subset \Phi^\times$ and $(S\Phi)\subset L^2[0,2\pi]\subset (S\Phi)^\times$.
The first one is related with the discrete basis $\{|m\rangle\}$ and in some sense is an abstract RHS, but the second one, related with the continuous basis $\{|\phi\rangle\}$, is obtained by means of a unitary map $S:|m\rangle\to e^{-im\phi}/\sqrt{2\pi}$ that allows to translate the topologies of the first RHS to the second one as well as all the relevant properties. 

Another interesting point to stress is the fact that RHS, from one side, and Lie algebras and Universal Enveloping Algebras, from the other,  are closely related. This means that starting from a  Lie algebra we can construct RHS that support its UIR and also that  generators and universal enveloping elements can be represented by continuous operators avoiding domain difficulties \cite{RO}.

%%%%%%%%%%%%%%%%%%%%%%%%%%%%%%%%%%%%%%%%%%%%%%%%
%%%%%%%%%%%%%%%%%%%%%%%%%%%%%%%%%%%%%%%%%%%%%%%%

\section*{Acknowledgements}
Partial financial support is acknowledged to the Spanish Junta de Castilla y Le\'on and FEDER (Project VA057U16) and MINECO (Project MTM2014-57129-C2-1-P).

%%%%%%%%%%%%%%%%%%%%%%%%%%%%%%%%%%%%%%%%%%%%%%%%
%%%%%%%% REFERENCES %%%%%%%%%%%%%%%%%%%%%%%%%%%%

\end{document}